# Current induced magnetization switching in Pt/Co/Cr structures with enhanced perpendicular magnetic anisotropy and spin-orbit torques

Baoshan Cui, Dong Li, Shiwei Chen, Jijun Yun, Yalu Zuo, Xiaobin Guo, Kai Wu, Xu Zhang, Yupei Wang, Dezheng Yang, Meizhen Gao, Li Xi[*]

Key Laboratory for Magnetism and Magnetic Materials of Ministry of Education & School of Physical Science and Technology, Lanzhou University, P. R. China

**Abstract**

Magnetic trilayers having large perpendicular magnetic anisotropy (PMA) and high spin-orbit torques (SOTs) efficiency are the key to fabricate nonvolatile magnetic memory and logic devices. In this work, PMA and SOTs are systematically studied in Pt/Co/Cr stacks as a function of Cr thickness. An enhanced perpendicular anisotropy field around 10189 ± 116 Oe is obtained and is related to the interface between Co and Cr layers. In addition, an effective spin Hall angle up to 0.19 is observed due to the improved antidamping-like torque by employing dissimilar metals Pt and Cr with opposite signs of spin Hall angles on opposite sides of Co layer. Finally, we observed a nearly linear dependence between spin Hall angle and longitudinal resistivity from their temperature dependent properties, suggesting that the spin Hall effect may arise from extrinsic skew scattering mechanism. Our results indicate that 3$d$ transition metal Cr with a large negative spin Hall angle could be used to engineer the interfaces of trilayers to enhance PMA and SOTs.

---

[*] Corresponding author. E-mail address: xili@lzu.edu.cn (Li Xi)

# I. INTRODUCTION

In recent years, current induced spin-orbit torques (SOTs) in trilayer structures, where ultrathin ferromagnets (FM) is sandwiched by a heavy metal (HM) and an oxide, have attracted abundant research interests for highly effective magnetization switching [1-3] and fast domain wall motion [4-10]. In such a SOT-based device, when an in-plane charge current ($J_e$) flows through HM with strong spin-orbit coupling (SOC) including 5$d$-metal Pt [2], Ta [11], W [12], and Hf [13], etc., it can be converted into a pure spin current ($J_s$). Then, $J_s$ injects into FM and generates a torque to act on magnetic moments. As a result, if the torque is sufficiently strong, the magnetization could be switched. It is well established that the SOT switching efficiency is directly related to the magnitude of spin Hall angle ($\theta_{SH}$). So, considerable efforts have been devoted to obtain a large $\theta_{SH}$. An enhancement of the SOT switching efficiency was reported in a Pt/Co/Ta structure [14], in which the spin Hall angles of Pt and Ta have opposite signs.

Moreover, as the perpendicular magnetic anisotropy field ($H_{an}^0$) is also one of the key parameters in SOT systems with perpendicular magnetic anisotropy (PMA), so far, considerable research efforts have also been devoted to obtain an enhanced $H_{an}^0$ to improve the thermal stability of spintronic devices [15, 16], such as engineering the interface quality [17, 18, 19], changing identified SOC materials [16], and utilizing different post processing methods [20]. In a word, all these reports with views are not only to increase the magnitude of $\theta_{SH}$, but also to enhance the PMA.

In this work, we explore the PMA and SOTs in Pt/Co/Cr structures as a function of Cr-thickness ($t$), where 3$d$ transition metal Cr with a large and negative $\theta_{SH}$ has been experimentally confirmed recently [21, 22]. Moreover, Pt and Cr were grown at an optimized sputtering condition in order to obtain large $\theta_{SH}$ as reported in W [23] and Pt [24]. In those conditions, the pure spin currents generated from Pt and Cr are enhanced and expected to work in concert to improve the SOT switching efficiency. Here, we use anomalous Hall effect (AHE) measurement setup [19] to characterize the PMA and SOT switching efficiency in Pt/Co/Cr stacks. Firstly, a remarkable enhancement of $H_{an}^0$ up to 10189 ± 116 Oe is obtained with $t$ = 5 nm. Secondly, SOT induced magnetization switching was achieved under a relatively small critical current density ($J_{crit}$) due to the improved SOT switching efficiency. Simultaneously, the relation between effective spin Hall angle ($\theta_{SH}^{eff}$) and longitudinal resistivity ($\rho_{xx}$) clarifies that the antidamping-like torque may mainly arise from extrinsic skew scattering mechanism in our systems. In addition, by measuring the extraordinary Hall resistance ($R_{Hall}$) under various applied direct currents ($I$), an obvious effect of Joule heating on magnetization switching was observed. The thermally activated domain wall (DW) nucleation and propagation mechanism is identified from the temperature dependent switching field ($H_{sw}$) measurements over a wide temperature ($T$) range from 150 to 475 K. Moreover, the PMA still remains with the $H_{sw}$ around 170 Oe when $T$ reaches to 475 K. Our findings suggest that Pt/Co/Cr systems could potentially be applied as the spintronics devices due to its significantly enhanced PMA and SOTs.

## II. EXPERIMENTAL DETAILS

The samples, with the structures Ta(3)/Pt(5)/Co(0.8)/Cr($t$)/Al(1) (thickness number in nanometer and $t = 1- 5$) were deposited on Corning glass substrate by direct current magnetron sputtering. The growth was carried out at a base pressure less than $5 \times 10^{-5}$ Pa. The relatively low sputtering power ($P$) with high Ar gas pressure ($p_{Ar}$) can increase the resistivity of Pt and Cr, which was identified to enhance $\theta_{SH}$ [23]. The optimized powers for 2 and 3 inches diameter Pt and Cr targets are 8 and 10 W, respectively, and the sputtering pressures for Pt and Cr are 0.53 and 0.99 Pa, respectively. We also note that the morphology and surface roughness of the sputtered films are known to depend on $P$ and $p_{Ar}$, atomic force microscopy (AFM) observation reveals that the sputtering condition employed here dose not significantly influence these parameters as shown in the supplementary materials. A Ta target was used to produce 3 nm seed layer and an Al target was used to produce 1 nm protective overcoat for our devices. All the structures were prepared at room temperature. The thickness of films was determined by the deposition time and sputtering rate, which was calibrated by X-ray reflectivity. The thin-film stacks were patterned into Hall bars by photolithography and Ar ion milling as shown in Fig. 1(a), with dimensions 6.8 × 180 $\mu m^2$. These devices were measured using a Keithley 220 current source, a Keithley 2000 multimeter for extraordinary Hall resistance measurements. A Keithley 6221 current source, and an Analog-Digital/Digital-Analog card for harmonic voltage measurements. For angular dependent planar Hall effect measurements, we used a physical-properties measurement system (PPMS).

## III. RESULTS AND DISCUSSIONS

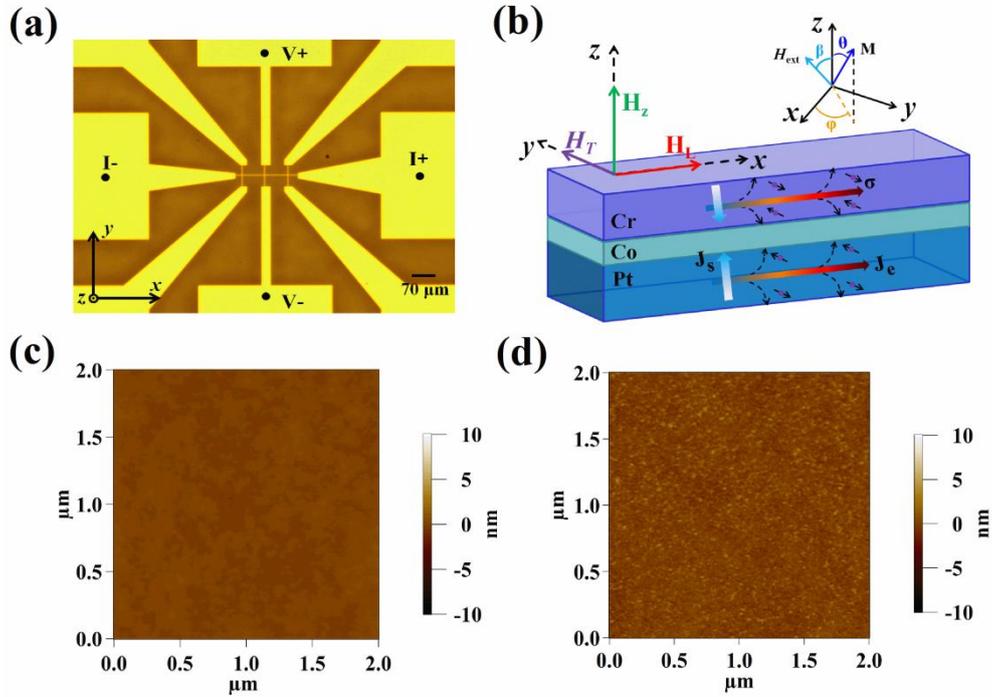

FIG. 1. (a) Optical photograph of the patterned Hall bar structure and the DC Hall resistance experimental configuration. (b) Schematic illustration of Pt/Co/Cr trilayers for magnetotransport measurement. 2×2 $\mu m^2$ AFM images of Pt (c) and Cr (d) films.

Optical photograph of the patterned Hall bar structure and the DC Hall resistance experimental configuration is shown in Fig.1 (a). Fig.1 (b) shows the schematic illustration of Pt/Co/Cr trilayers for magnetotransport measurement. When a charge current ($J_e$) passing through Pt and Cr along the length of Hall bar (x-axis), a spin current ($J_s$) will be generated perpendicular to the Pt and Cr layers (along the z-axis) via spin Hall effect (SHE). The 2 × 2 $\mu m^2$ AFM images of Pt and Cr layers with the thickness around 30 nm are shown in Figs. 1(c) and (d), respectively. Whereas, a small roughness with root-mean square roughness around 0.20 and 0.62 nm is obtained for Pt and Cr, respectively, indicating that the high-quality films can be obtained in our experimental conditions. The saturation magnetizations ($M_s$) are around 736, 752, 1104, 921, and 943 emu/cc for the samples with $t$ = 1, 2, 3, 4, and 5 nm, respectively, with the same nominal Co thickness 0.8 nm. The measured $M_s$ are inconsistent with each other and can be ascribed to the Co thickness variation during sputtering. The calculated Co-thickness from $M_s$ is around 0.8 ± 0.09 nm, which is quite reasonable for our home made sputtering machine.

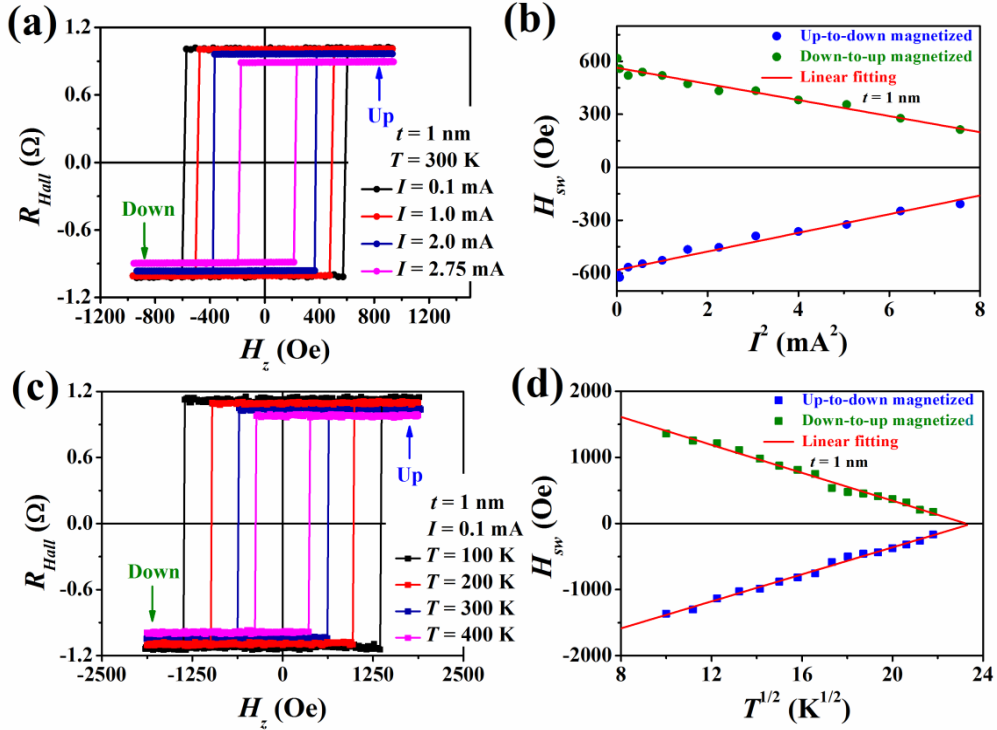

FIG. 2. (a) $R_{Hall}$-$H_z$ loops for various DC currents. (b) The linear fitting with the relationship between $H_{sw}$ and $I^2$. (c) Square-like $R_{Hall}$-$H_z$ loops at different temperatures. (d) The linear fitting with the relationship between $H_{sw}$ and $T^{1/2}$.

Fig. 2 shows the magnetic properties of Pt(5)/Co(0.8)/Cr(1) characterized by AHE measurement. As one knows, $R_{Hall}$ is proportional to the perpendicular component of magnetization ($M_z$) of the Co layer, which can thus be used to identify the direction of magnetization with $M_z^{up}$ ($M_z^{down}$) corresponding to $R_{Hall} > 0$ ($R_{Hall} < 0$) as labeled in Figs. 2(a) and (c). Fig. 2(a) illustrates the $R_{Hall}$ as a function of perpendicular magnetic field ($H_z$) under a series of *I*. The square-like shaped loops confirm the presence of PMA. Moreover, the switching field ($H_{sw}$) extracted from $R_{Hall}$-$H_z$ loops exhibits a linear decrease with $I^2$ increasing as shown in Fig. 2(b), which reveals that the Joule heating effect may play a key role in $H_{sw}$.

Fig. 2(c) shows the representative $R_{Hall}$-$H_z$ loops over a large temperature range from 100 to 475K for the sample with $t = 1$ nm. In order to decrease the influence of the Joule heating effect and SOTs, a quite small current with $I = 0.1$ mA was employed during the measurements. One can clearly see the square-like loops even for samples at high temperature, indicating that a robust PMA is sustained over the whole temperature rang. It must also be mentioned that the $H_{sw}$ and $R_0$ (the Hall resistance with $H_z = 0$) both decrease significantly as temperature increases. Additionally, as can be seen from Fig. 2(d), $H_{sw}$ displays a linear dependence on the square root of $T$. The strong temperature dependence in $H_{sw}$ is an indication of the mechanism of thermally activated domain wall (DW) motion, it can be expressed as [25, 26]

$$H_{sw}(T) = H_{sw}(0)\left[1 - aT^{\frac{1}{2}}\right] \quad (1)$$

where, $H_{sw}(0)$ is the switching field at 0 K, and the constant $a$ depends on the activation energy of DW motion. The solid lines in the Fig. 2(d) are the fitting of the data points based on Eq. (1). We can clearly see that $H_{sw}$ points are fitted well by Eq. (1). From the fitting, one can see that $H_{sw}$ will decrease to zero at the temperature around 542 K. Moreover, a large $H_{sw}$ was obtained even at high temperature (e.g. $H_{sw}^0 = 170$ Oe at $T = 475$ K), indicating that Pt/Co/Cr systems have a quite good thermal stability, which is fairly important for applications.

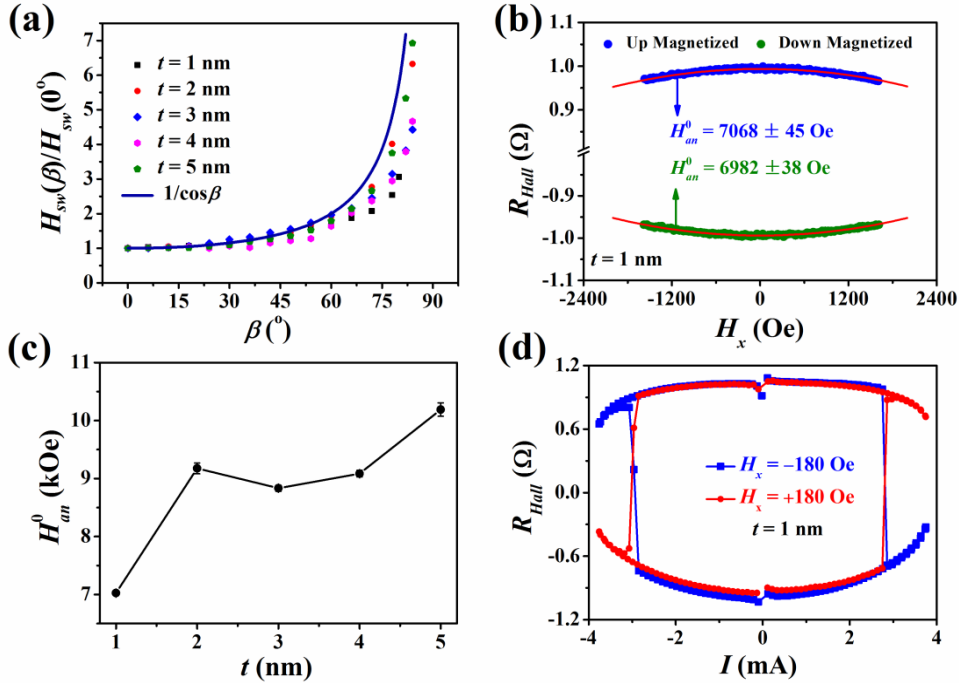

FIG. 3. (a) $H_{sw}(\beta)/H_{sw}(0^o)$ as a function of $\beta$ together with an inverse $\cos\beta$ curve. (b) $H_x$ dependence of $R_{Hall}$ under a low direct current ($I_{DC} = +0.5$ mA). The solid lines are fitting curves.

(c) Thickness dependence of $H_{an}^0$. (d) Current-induced magnetization switching at room temperature in the presence of a small fixed magnetic field $H_x = \pm 180$ Oe.

To further confirm the magnetization reversal mechanism, we also summarize $H_{sw}$ against the angle $β$ ($β$ is the angle between the $H_{ext}$ and z-axis in x-z plane) in Fig. 3(a) obtained from the angular dependent $R_{Hall}$-$H_{ext}$ loops which are measured at $I = 0.1$ mA to reduce the spin Hall torque and Joule heating effect. Kondorsky firstly pointed out a magnetization switching mechanism model for uniaxial highly anisotropic system with a depinning of 180° domain wall or nucleation and growth of reverse domains using the angular dependent $H_{sw}$ [27]. In this model, the magnetization will reverse when the component of the $H_{ext}$ parallel to z-axis overcomes the nucleation field or depinning field ($H_d$) established at $β = 0°$, i.e. [27-29],

$$H_{sw}(β) = H_{sw}(0°)/\cos β \qquad (2)$$

Obviously, the measured data points exactly follow an inverse $\cos β$ curve (shown as mazarine solid line) when $β < 30°$. The agreement of the experimental and theoretical prediction indicates the dominated depinning or nucleation and growth of reverse domains process for magnetization switching in a small angle range [29]. While, the experimental points slightly deviate from the inverse $\cos β$ curve with large $β$, indicating that the coherent rotation process gradually contributes to the magnetization switching since the projection of $H_{ext}$ along the x-axis gradually increases with $β$ [28]. Furthermore, for the sample with $t = 3$ nm, the experimental points just deviate from the inverse $\cos β$ curve slightly even in the highest angle $β$ of 60°, implying a quite strong perpendicular magnetic anisotropy for this sample. In the same way, the $H_{sw}$ remarkably deviates from the $1/\cos β$ dependence for other samples corresponding to relatively weak PMA. Actually, the nucleation and fast sweeping of reversed domain wall process was observed for most of Hall bars with the quite fast change of the contrast for domain observation, while the depinning process was also observed for some sample using a Kerr microscopy for domain wall observation as shown in the supplemented videos.

From the above magnetization switching mechanism, one can see that the switching only occurs when the reversed magnetic field along the z direction reaches the $H_{sw}$. Keeping this in mind, we now analysis the current induced magnetization switching by SOTs with the assistant of an in- plane x-directional magnetic field ($H_x$). We first consider the equilibrium equation of the magnetization using a simple macrospin model considering the external field torque, the anisotropy torque and the spin orbital torque [1, 30, 31]. When $J_e$ flowing along $+\hat{x}$ direction, the SHE induced $J_s$ is generated perpendicular to the Co layer (along the $+\hat{z}$ direction from Pt layer and along the $-\hat{z}$ direction from Cr layer) as shown in Fig. 1(b). Meanwhile, $H_x$ was applied to the sample along x-axis. As a result, the direction of magnetization *M* of Co layer is controlled by $τ_{ext}$ (the torque generated by $H_x$, i.e. $\vec{τ}_{ext} = \vec{M} \times \vec{H}_x$, oriented along $+\hat{y}$) and the SHE induced spin orbital torque (SOT, i.e., $\vec{τ}_{ST} = τ_{ST}^0 (\vec{M} \times \vec{σ} \times \vec{M}) = \left(\frac{\hbar}{2eM_s t_{Co}} J_s\right)(\vec{M} \times \vec{σ} \times \vec{M})$, oriented along $+\hat{y}$). In addition, we must take into account the torque due to the perpendicular anisotropy field, i.e., $\vec{\tau}_{an} = \vec{M} \times \vec{H}_{an}$ parallel to $-\hat{y}$. In this case, all torques ($\vec{\tau}_{ST}, \vec{\tau}_{ext}, \vec{\tau}_{an}$) are collinear in y-axis and the equilibrium condition for *M* is

$$\tau_{tot} = \hat{y} \cdot (\vec{\tau}_{ST} + \vec{\tau}_{ext} + \vec{\tau}_{an})$$
$$= \tau_{ST}^0 + H_x \cos\theta - H_{an}^0 \sin\theta \cos\theta = 0 \quad (3)$$

when $J_e$ is quite small, thus the current generated SOTs can be ignored, i.e., $\tau_{ST}^0 \approx 0$, and Eq. (3) is reduced to

$$H_x - H_{an}^0 \sin\theta = 0 \quad (4)$$

As we know, the $R_{Hall}$ is proportional to the $M_z$. Thus, the $\theta$ can be determined by

$$\cos\theta = R_{Hall}(\theta) / R_0 \quad (5)$$

By solving the simultaneous equations using the combination of Eq. (4) and (5), one can obtain

$$R_{Hall}(\theta) = R_0 \cdot \cos\left(\arcsin \frac{H_x}{H_{an}^0}\right) \quad (6)$$

Fig. 3(b) shows the $H_x$ dependence of $R_{Hall}$ under a low-direct current (*I* = +0.5 mA). We can obtain the $H_{an}^0$ by fitting the data using Eq. (6) as shown in Fig. 3(c). As expected, a larger $H_{an}^0 \approx$ 10189 ± 116 Oe with *t* = 5 nm is obtained, $H_{an}^0$ of several SOT devices with PMA are shown in the supplement materials. As far as we know, $H_{an}^0$ obtained in this work is the largest value compared to other PMA structures, such as Pt/Co/Pt [32], Pt/Co/Ta [14], Ta/CoFeB/MgO [30].

Although, a quite strong PMA is obtained in our Pt/Co/Cr stacks, the current-induced SOTs could overcome the anisotropy barrier to achieve the switching of magnetization according to Eq. (3). However, as discussed above, the magnetization reversal mechanism for our devices is governed by the nucleation and/or depinning mechanism. Thus, the effective magnetic field along the *z* direction from SOT only needs to reach the $H_{sw}$, which is much smaller than the anisotropy field, to achieve the magnetization switching. Exhilaratingly, current-induced magnetization switching (CIMS) not only can be achieved with $H_x$ = 180 Oe, but also under relatively small critical current density ~4.5 × 10$^6$ A/cm$^2$ as depicted in Fig. 3(d). It can be ascribed to the improved SOTs in Pt/Co/Cr systems. Moreover, the directions of current and $H_x$ determine the polarity of the switching, which is consistent with the model of the SHE switching [11].

The current-induced SOT effective fields were measured by the harmonic Hall voltage measurement technique. The measurement diagram is depicted in Fig. 1(b). A sinusoidal alternate current (AC) with a frequency of 133 Hz along *x*-axis was passed through the Hall bars and the first ($V^\omega$) and second ($V^{2\omega}$) harmonic Hall voltage were collected using a Analog-Digital/Digital-Analog card by the frequency-spectra analysis. The current-induced longitudinal antidamping-like field ($H^{DL}$) and the transverse field-like field ($H^{FL}$) can be obtain by the $H_{ext}$ applied along the *x* (longitudinal-field $H_L$) and *y* (transverse-field $H_T$) directions, respectively. Then, the AC-induced $H^{DL(FL)}$ along a given in-plane field direction can be determined as [33]

$$\Delta H^{DL(FL)} = \frac{H^{DL(FL)} \pm 2\xi H^{FL(DL)}}{1-4\xi^2}$$

$$H^{DL(FL)} = -2 \frac{\partial V^{2\omega}}{\partial H_{L(T)}} \Big/ \frac{\partial^2 V^{\omega}}{\partial H_{L(T)}^2}$$

$$\xi = \Delta R_P / \Delta R_A \qquad (7)$$

where, $\xi$ is the ratio of planar Hall resistance ($\Delta R_P$) over anomalous Hall resistance ($\Delta R_A$). $\Delta H^{DL(FL)}$ is the corrected antidamping-like (field-like) field considering $\xi$. Figs. 4(a) and (b) respectively show $V^{\omega}$ as a function of $H_L$ and $H_T$ with $t = 1$ nm sample. The insets of each figure represent $V^{2\omega}$ as a function of $H_L$ and $H_T$. Furthermore, in order to obtain $H^{DL}$ and $H^{FL}$, the first and second harmonic curves were fitted using the quadratic and linear fitting function, respectively, based on Eq. (7). The best fitting (red solid lines) are shown in Figs. 4(a) and (b). The obtained $H^{DL}$ and $H^{FL}$ are plotted with respect to the amplitude ($I_0$) of AC in Fig. 4(c). One can see that $H^{DL}$ linearly increases with $I_0$. It indicates that the nonlinear effects are negligible at lower AC, and the slope of the fitted curves gives $\beta^{DL} = H^{DL}/J_e$, which are exhibited in Fig. 4(d). Interestingly, the $H^{FL}$ is quite small even on a large $I_0$.

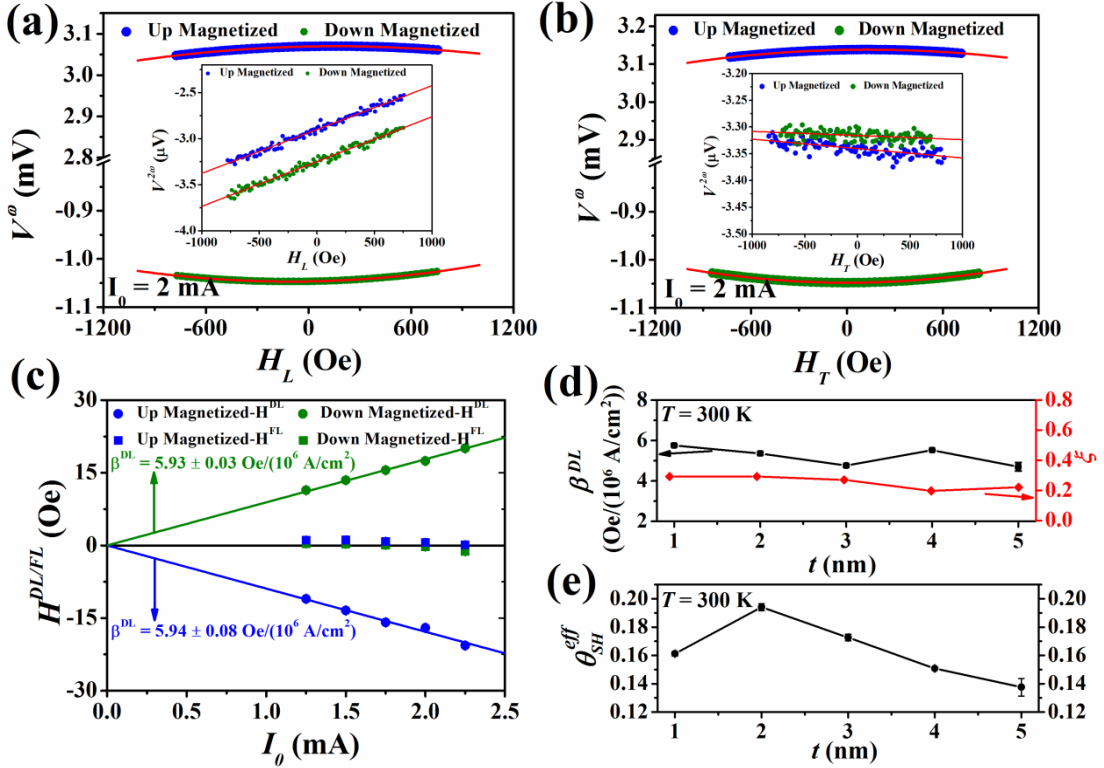

FIG. 4. (a)-(b) $V^{\omega}$ versus $H_L$ and $H_T$ with $I_0 = 2$ mA. The inset in each figure shows the $V^{2\omega}$ as a function of $H_L$ and $H_T$. The red lines stand for the quadratic and linear fitting. (c) $H^{DL/FL}$ against the amplitude of the sinusoidal current. (d) Thickness dependence of $H^{DL}$ for per unit current density ($\beta^{DL} = H^{DL}/J_e$) and the ratio of planar Hall resistance to the anomalous Hall resistance ($\xi = \Delta R_P/\Delta R_A$). (e) Thickness dependence of $\theta_{SH}^{eff}$.

Since the measured Hall voltages simultaneously include the contributions from the planar Hall effect (PHE) and AHE, the $R_{Hall}$ can be expressed by [33],

$$R_{Hall} = \frac{1}{2}\Delta R_A \cos\theta + \frac{1}{2}\Delta R_P \sin^2\theta \sin 2\varphi \tag{8}$$

where the first and second terms respectively represent AHE and PHE. $\theta$ is polar angle, $\varphi$ is azimuthal angle in a spherical coordinate system as shown in Fig. 1(b). $\Delta R_P$ was obtained by measuring $R_{Hall}$ as a function of $\varphi$ with $\theta = 90°$ and $\Delta R_A$ was obtained from $R_{Hall}$-$H_z$ loops. A series of $\xi$ were calculated as shown in Fig. 4(b). Large values of $\xi$ indicate that the influence of PHE to the SOT fields is more significant and should not be neglected in our systems.

Based on Eqs. (7), the $\Delta H^{DL}$ were obtained, then, we quantitatively calculated the $\theta_{SH}^{eff}$ to compare the strength of the antidamping-like torque in the present systems using the following equation [30],

$$\theta_{SH}^{eff} = \frac{J_s}{J_e} = \frac{2eM_s t_{FM}}{\hbar} \cdot \frac{\Delta H^{DL}}{J_e} \tag{9}$$

where $e$ is the elementary charge, $\hbar$ is reduced Planck constant. $t_{FM}$ is the thickness of ferromagnetic layer, and $M_s$ is saturation magnetization. Assuming that the current flows equably in our devices, the $\theta_{SH}^{eff}$ were calculated from Eq. (9) as shown in Fig. 4(e). We can see that $\theta_{SH}^{eff}$ are quite large comparing with the spin Hall angle of Pt ($\theta_{SH}^{Pt} \sim 0.07$) [1] in our devices. Especially, $\theta_{SH}^{eff}$ of up to 0.19 with $t = 2$ nm is obtained. Those results are ascribed to the SOTs from each interface working in concert to enhance the total effective torque causing by employing the dissimilar metals Pt and Cr with opposite signs of spin Hall angles on opposite sides of Co layer. On the other hand, the enhanced $\theta_{SH}^{eff}$ may also be attributed to the increased resistivity of Pt and Cr under the optimized sputtering conditions.

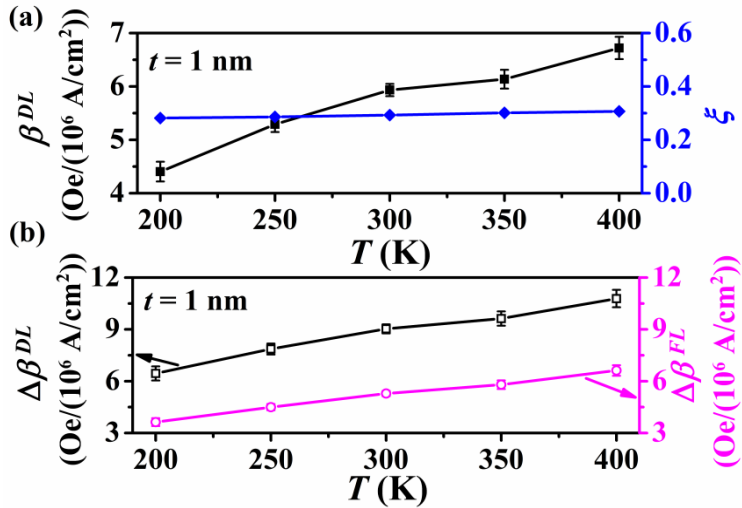

FIG. 5. (a) Temperature dependence of $H^{DL}$ for per unit current density ($\beta^{DL} = H^{DL}/J_e$) and $\xi$. (b) Temperature dependence of corrected antidamping-like (field-like) effective field for per unit current density ($\Delta\beta^{DL/FL} = \Delta H^{DL/FL}/J_e$).

Finally, we investigate the temperature dependence of SOTs. Similarly, the harmonic Hall voltage measurement technique was used to quantify the current-induced SOT effective fields over a large temperature range from 200 to 400 K. Fig. 5(a) shows the temperature dependence of $\beta^{DL}$ and $\xi$ for the sample with $t = 1$ nm. One can see that $\beta^{DL}$ increases with temperature increasing, but $\xi$ shows a weak temperature dependence. Notably, although the $H^{FL}$ is negligible, but the $\Delta H^{FL}$ can not be ignored due to the large $H^{DL}$ according to Eq. (7). Fig. 5(b) depicts the corrected antidamping-like and field-like effective field for per unit current density ($\Delta \beta^{DL/FL} = \Delta H^{DL/FL}/J_e$) as a function of $T$, both types of torques exhibit nearly linear dependence on $T$. Among them, the temperature dependence of $\Delta \beta^{DL}$ is similar to that in CuAu/FeNi/Ti stacks [34], in which extrinsic SHE is the dominant mechanism rather than intrinsic SHE. In other words, the extrinsic SHE i.e. the scattering events are increasing with the temperature increasing, however, intrinsic SHE is not significantly affected. We attributed the temperature dependence of $\Delta \beta^{DL}$ to the similar mechanism, which will be further discussed below.

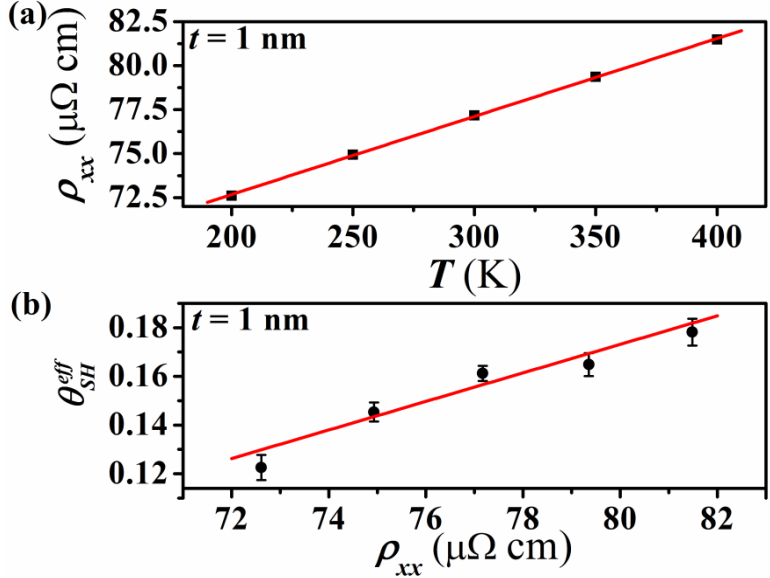

FIG. 6.(a) $\rho_{xx}$ as a function of $T$ from 200 to 400 K. (b) Relation between $\theta_{SH}^{eff}$ and $\rho_{xx}$.

To further gain a deeper understanding of the dependence of the antidamping-like torque on temperature, we studied the relationship between $\theta_{SH}^{eff}$ and longitudinal resistivity ($\rho_{xx}$) over a temperature range from 200 to 400 K. Theoretically, anomalous Hall resistivity ($\rho_{AH}$) in ferromagnetic materials as the sum of the skew-scattering term ($a\rho_{xx}$) and the extrinsic side jump or intrinsic mechanism term ($b\rho_{xx}^2$), viz. [35],

$$\rho_{AH} = a\rho_{xx} + b\rho_{xx}^2 \tag{10}$$

In metals with low resistivity, the scaling relation of spin Hall angle $\theta_{SH} \propto \rho_{xx}$ is most likely. The typically weak $T$ dependence of $\rho_{xx}$ is presented in Fig. 6(a). A small variation ~ 12.2% of $\rho_{xx}$ is obtained ranging from 200 to 400 K. It is evident that temperature dependent phonon electron scattering is not the main source of the change in $\rho_{xx}$. Instead, scattering mechanism may play the dominant role. Fig. 6(b) shows the relation between $\theta_{SH}^{eff}$ and $\rho_{xx}$. The linear correlation may be the best evidence to illustrate that the skew scattering may be the dominant source for the SHE in our samples [30].

The temperature dependence of corrected field-like torque ($\Delta\beta^{FL}$) also deserves some discussion. As shown in Fig. 5(b), $\Delta\beta^{FL}$ as the scenario of interfacial Rashba torque increases with temperature, the increase could be attributed to an increase in bulk resistance with temperature increasing, as a result, increasing the current flowing through the interface. Thereby, the enhancement of corrected field-like torque can be observed. We emphasize that this explanation remains speculative and requires further experiments to confirm.

## IV. CONCLUSIONS

In summary, we have performed a comprehensive study on the perpendicular magnetic anisotropy, magnetization switching mechanism and spin-orbit torques of perpendicularly magnetized Pt/Co/Cr trilayers. The obtained perpendicular anisotropy field reaches up to 10189 ± 116 Oe, which is the largest value observed to date in metallic sandwich systems. We ascribe the enhanced perpendicular anisotropy field to the interface coupling between Co and Cr layers. In addition, a spin-orbit torque induced magnetization switching is achieved under relatively small critical current density in the order of $10^6$ A/cm$^2$ due to the improved SOTs in Pt/Co/Cr systems. The improved SOTs reveal a large effective spin Hall angle around 0.19 by employing the dissimilar metals Pt and Cr with opposite signs of spin Hall angles. Moreover, the switching field extracted from $R_{Hall}$-$H_z$ loops exhibit a strong $I^2$ ($I$: applied-direct current) dependence, which may be attributed to the vital role of Joule heating effect. Importantly, the antidamping-like torque and field-like torque are corrected by the ratio of planar Hall resistance to the anomalous Hall resistance, both types corrected torques are prominent and displaying a nearly linear increase with temperature. The temperatures dependence of antidamping-like and field-like torque may, respectively, arise from extrinsic skew scattering and enhanced Rashba effect. Our results indicate that 3$d$ transition metal Cr with a large negative spin Hall angle could be used to engineer the interfaces of trilayers to enhance PMA and SOTs, which may be benefited to future spintronic devices.

## ACKNOWLEDGMENTS

This work was supported by the Program for Changjiang Scholars and Innovative Research Team in University PCSIRT (No. IRT16R35), the National Natural Science Foundation of China (No. 51671098), the Natural Science Foundation of Gansu Province (No. 145RJZA154) and the Fundamental Research Funds for the Central Universities (lzujbky-2015-122).

# References

[1] L. Liu, O. J. Lee, T. J. Gudmundsen, D. C. Ralph, and R. A. Buhrman, Current-Induced Switching of Perpendicularly Magnetized Magnetic Layers Using Spin Torque from the Spin Hall Effect, Phys. Rev. Lett. **109**, 096602 (2012).

[2] I. M. Miron, K. Garello, G. Gaudin, P.-J. Zermatten, M. V. Costache, S. Auffret, S. Bandiera, B. Rodmacq, A. Schuhl, and P. Gambardella, Perpendicular switching of a single ferromagnetic layer induced by in-plane current injection, Nature **476**, 189 (2011).

[3] M. Jamali, K. Narayanapillai, X. P. Qiu, L. M. Loong, A. Manchon, and H. Yang, Spin-Orbit Torques in Co/Pd Multilayer Nanowires, Phys. Rev. Lett. **111**, 246602 (2013).

[4] S. Emori, U. Bauer, S.-M. Ahn, E. Martinez, and G. S. D. Beach, Current-driven dynamics of chiral ferromagnetic domain walls, Nat. Mater. **12**, 611 (2013).

[5] P. P. J. Haazen, E. Murè, J. H. Franken, R. Lavrijsen, H. J. M. Swagten, and B. Koopmans, Domain wall depinning governed by the spin Hall effect, Nat. Mater. **12**, 299 (2013).

[6] T. A. Moore, I. M. Miron, G. Gaudin, G. Serret, S. Auffret, B. Rodmacq, A. Schuhl, S. Pizzini, J. Vogel, and M. Bonfim, High domain wall velocities induced by current in ultrathin Pt/Co/AlOx wires with perpendicular magnetic anisotropy, Appl. Phys. Lett. **93**, 262504 (2008).

[7] I. M. Miron, T. Moore, H. Szambolics, L. D. Buda-Prejbeanu, S. Auffret, B. Rodmacq, S. Pizzini, J. Vogel, M. Bonfim, A. chuhl, and G. Gandi, Fast current-induced domain-wall motion controlled by the Rashba effect, Nat. Mater. **10**, 419 (2011).

[8] D. Chiba, M. Kawaguchi, S. Fukami, N. Ishiwata, K. Shimamura, K. Kobayashi, and T. Ono, Electric-field control of magnetic domain-wall velocity in ultrathin cobalt with perpendicular magnetization, Nat. Commun. **3**, 888 (2012).

[9] K.-S. Ryu, L. Thomas, S.-H. Yang, and S. Parkin, Chiral spin torque at magnetic domain walls, Nat. Nanotechnol. **8**, 527 (2013).

[10] S.-H. Yang, K.-S. Ryu, and S. Parkin, Domain-wall velocities of up to 750 m s$^{-1}$ driven by exchange-coupling torque in synthetic antiferromagnets, Nat. Nanotechnol. **10**, 221 (2015).

[11] L. Liu, C.-F. Pai, Y. Li, H. W. Tseng, D. C. Ralph, and R. A. Buhrman, Spin-Torque Switching with the Giant Spin Hall Effect of Tantalum, Science **336**, 555 (2012).

[12] C.-F. Pai, L. Liu, Y. Li, H. W. Tseng, D. C. Ralph, and R. A. Buhrman, Spin transfer torque devices utilizing the giant spin Hall effect of tungsten, Appl. Phys. Lett. **101**, 122404 (2012).

[13] R. Ramaswamy, X. Qiu, T. Dutta, S. D. Pollard, and H. Yang, Hf thickness dependence of spin-orbit torques in Hf/CoFeB/MgO heterostructures, Appl. Phys. Lett. **108**, 202406 (2016).

[14] S. Woo, M. Mann, A. J. Tan, L. Caretta, and G. S. D. Beach, Enhanced spin-orbit torques in Pt/Co/Ta heterostructures, Appl. Phys. Lett. **105**, 212404 (2014).

[15] J. Sinha, M. Hayashi, A. J. Kellock, S. Fukami, M. Yamanouchi, H. Sato, S. Ikeda, S. Mitani, S. H. Yang, S. S. P. Parkin, and H. Ohno, Enhanced interface perpendicular magnetic anisotropy in Ta|CoFeB|MgO using nitrogen doped Ta underlayers, Appl. Phys. Lett. **102**, 242405 (2013).

[16] C.-F. Pai, M. H. Nguyen, C. Belvin, L. H. Vilela-Leao, D. C. Ralph, and R. A. Buhrman, Enhancement of perpendicular magnetic anisotropy and transmission of spin-Hall-effect-induced spin currents by a Hf spacer layer in W/Hf/CoFeB/MgO layer structures, Appl. Phys. Lett. **104**, 082407 (2014).

[17] S.-L. Jiang, X. Chen, X.-J. Li, K. Yang, J.-Y. Zhang, G. Yang, Y.-W. Liu, J.-H. Lu, D. W. Wang, J. Teng, and G.-H. Yu, Anomalous Hall effect engineering via interface modification in Co/Pt multilayers, Appl. Phys. Lett. **107**, 112404 (2015).

[18] C.-F. Pai, Y. Ou, L. H. Vilela-Leao, D. C. Ralph, and R. A. Buhrman, Dependence of the efficiency of spin Hall torque on the transparency of Pt/ferromagnetic layer interfaces, Phys. Rev. B **92**, 064426 (2015).

[19] D. Li, B. S. Cui, T. Wang, J. J. Yun, X. B. Guo, K. Wu, Y. L. Zuo, J. B. Wang, D. Z.Yang, and L. Xi, Effect of inserting a non-metal C layer on the spin-orbit torque induced magnetization switching in Pt/Co/Ta structures with perpendicular magnetic anisotropy, Appl. Phys. Lett. **110**, 132407 (2017)

[20] W.-G. Wang, S Hageman, M. Li, S. Huang, X. Kou, X. Fan, J. Q. Xiao, and C. L. Chien, Rapid thermal annealing study of magnetoresistance and perpendicular anisotropy in magnetic tunnel junctions based on MgO and CoFeB, Appl. Phys. Lett. **99**, 102502 (2011).

[21] C. Du, H. Wang, F. Yang, and P. C. Hammel, Systematic variation of spin-orbit coupling with *d*-orbital filling: Large inverse spin Hall effect in 3*d* transition metals, Phys. Rev. B **90**, 140407(R) (2014).

[22] D. Qu, S. Y. Huang, and C. L. Chien, Inverse spin Hall effect in Cr: Independence of antiferromagnetic ordering, Phys. Rev. B **92**, 020418(R) (2015).

[23] C. Zhang, S. Fukami, K. Watanabe, A. Ohkawara, S. DuttaGupta, H. Sato, F. Matsukura, and H. Ohno, Critical role of W deposition condition on spin-orbit torque induced magnetization switching in nanoscale W/CoFeB/MgO, Appl. Phys. Lett. **109**, 192405 (2016).

[24] E. Sagasta, Y. Omori, M. Isasa, M. Gradhand, L. E. Hueso, Y. Niimi, Y. Otani, and F. Casanova, Tuning the spin Hall effect of Pt from the moderately dirty to the superclean regime, Phys. Rev. B **94**, 060412(R) (2016).

[25] W. C. Nunes, W. S. D. Folly, J. P. Sinnecker, and M. A. Novak, Temperature dependence of the coercive field in single-domain particle systems, Phys. Rev. B **70**, 014419 (2004).

[26] M.-C. Tsai, C.-W. Cheng, C. C. Tsai, and G. Chern, The intrinsic temperature dependence and the origin of the crossover of the coercivity in perpendicular MgO/CoFeB/Ta structures, J. Appl. Phys. **113**, 17C118 (2013).

[27] F. Schumacher, On the modification of the Kondorsky function, J. Appl. Phys. **70**, 3184 (1991).

[28] O. J. Lee, L. Q. Liu, C. F. Pai, Y. Li, H. W. Tseng, P. G. Gowtham, J. P. Park, D. C. Ralph, and R. A. Buhrman, Central role of domain wall depinning for perpendicular magnetization switching driven by spin torque from the spin Hall effect, Phys. Rev. B **89**, 024418 (2014).

[29] A. Singh, V. Neu, S. Fähler, K. Nenkov, L. Schultz, and B. Holzapfel, Mechanism of coercivity in epitaxial SmCo5 thin films, Phys. Rev. B **77**, 104443 (2008).

[30] Q. Hao, and G. Xiao, Giant spin Hall effect and magnetotransport in a Ta/CoFeB/MgO layered structure: A temperature dependence study, Phys. Rev. B **91**, 224413 (2015).

[31] Q. Hao, and G. Xiao, Giant Spin Hall Effect and Switching Induced by Spin-Transfer Torque in a W/Co$_{40}$Fe$_{40}$B$_{20}$/MgO Structure with Perpendicular Magnetic Anisotropy, Phys. Rev. Appl. **3**, 034009 (2015).

[32] P. J. Metaxas, J. P. Jamet, A. Mougin, M. Cormier, J. Ferré, V. Baltz, B. Rodmacq, B. Dieny, and R. L. Stamps, Creep and Flow Regimes of Magnetic Domain-Wall Motion in Ultrathin Pt∕Co∕Pt Films with Perpendicular Anisotropy, Phys. Rev. Lett. **99**, 217208 (2007).

[33] M. Hayashi, J. Kim, M. Yamanouchi, and Hideo Ohno, Quantitative characterization of the spin-orbit torque using harmonic Hall voltage measurements, Phys. Rev. B **89**, 144425 (2014)

[34] Y. Wen, J. Wu, P. Li, Q. Zhang, Y. L. Zhao, A. Manchon, J. Q. Xiao, and X. X. Zhang, Temperature dependence of spin-orbit torques in Cu-Au alloys, Phys. Rev. B **95**, 104403 (2017)

[35] N. Nagaosa, J. Sinova, S. Onoda, A. H. MacDonald, and N. P. Ong, Anomalous Hall effect, Rev. Mod. Phys. **82**, 1539 (2010).